# DIPOLE WAKEFIELD SUPPRESSION IN HIGH PHASE ADVANCE DETUNED LINEAR ACCELERATORS FOR THE JLC/NLC DESIGNED TO MINIMISE ELECTRICAL BREAKDOWN AND CUMULATIVE BBU*

R.M. Jones*, SLAC; N.M. Kroll, UCSD & SLAC; T. Higo, KEK, Z. Li, R.H. Miller,
T.O. Raubenheimer, and J.W. Wang,
Stanford Linear Accelerator Center, Stanford, CA 94309

## Abstract

Recent experiments at SLAC [1,2] and CERN [3] have revealed evidence of significant deformation in the form of "pitting" of the cells of the 1.8m series of structures DDS/RDDS (Damped Detuned Structure/Rounded Damped Detuned Structure). This pitting occurs in the high group velocity ($v_g/c = 0.012$) end of the accelerating structure and little evidence of breakdown has been found in the lower group velocity end of the structure. Additional, albeit preliminary experimental evidence, suggests that shorter and lower group velocity structures have reduced breakdown events with increasing accelerating field strengths. Two designs are presented here, firstly a 90cm structure consisting of 83 cells with an initial $v_g/c = 0.0506$ (known as H90VG5) and secondly, an even shorter structure of length 60cm consisting of 55 cells with an initial $v_g/c = 0.03$ (known as H60VG3). The feasibility of using these structures to accelerate a charged beam over 10km is investigated. The particular issue focussed upon is *suppression of the dipole wakefields* via detuning of the cell frequencies and by locally damping individual cells in order to avoid BBU (Beam Break Up). Results are presented on beam-induced dipole wakefields and on the beam dynamics encountered on tracking the progress of the beam through several thousand accelerating structures.

*Paper presented at the 2001 Particle Accelerator Conference*
*Chicago, USA*
*June 18$^{th}$ –June 22$^{nd}$, 2001*

___________________
*Supported under U.S. Department Of Energy contract DE-AC03-76SF00515
†Supported under U.S. Department Of Energy grant DE-FG03-93ER40

# DIPOLE WAKEFIELD SUPPRESSION IN HIGH PHASE ADVANCE DETUNED LINEAR ACCELERATORS FOR THE JLC/NLC DESIGNED TO MINIMISE ELECTRICAL BREAKDOWN AND CUMULATIVE BBU*

R.M. Jones*, SLAC; N.M. Kroll, UCSD & SLAC; T. Higo, KEK, Z. Li, R.H. Miller,
T.O. Raubenheimer, and J.W. Wang SLAC

**Abstract**

Recent experiments at SLAC [1,2] and CERN [3] have revealed evidence of significant deformation in the form of "pitting" of the cells of the 1.8m series of structures DDS/RDDS (Damped Detuned Structure/Rounded Damped Detuned Structure). This pitting occurs in the high group velocity ($v_g/c = 0.012$) end of the accelerating structure and little evidence of breakdown has been found in the lower group velocity end of the structure. Additional, albeit preliminary experimental evidence, suggests that shorter and lower group velocity structures have reduced breakdown events with increasing accelerating field strengths. Two designs are presented here, firstly a 90cm structure consisting of 83 cells with an initial $v_g/c = 0.0506$ (known as H90VG5) and secondly, an even shorter structure of length 60cm consisting of 55 cells with an initial $v_g/c = 0.03$ (known as H60VG3). The feasibility of using these structures to accelerate a charged beam over 10km is investigated. The particular issue focussed upon is *suppression of the dipole wakefields* via detuning of the cell frequencies and by locally damping individual cells in order to avoid BBU (Beam Break Up). Results are presented on beam-induced dipole wakefields and on the beam dynamics encountered on tracking the progress of the beam through several thousand accelerating structures.

## 1. INTRODUCTION

Recent experimental results have revealed significant damage to the cells of the DDS/RDDS series of accelerating structures due to RF breakdown occurring. A concerted both theoretical and experimental effort has been mounted at SLAC and KEK in order to understand why it occurs in our particular structures and to prevent electrical breakdown occurring in further structures.

Application of a simple circuit model [4] allows an understanding of the basic processes that may be involved in the breakdown and this has been applied to our travelling wave structures to obtain the power absorbed in the breakdown [5] (modelled as a single resistor, $R_A$ at a localised cell):

$$P_{abs} \simeq \frac{R_A}{(R/Q)^2} \frac{v_g^2}{\omega/c} \frac{4\sin\phi \,(grad)^2}{\phi\sin\phi + 2v_g\cos\phi} \quad (1.1)$$

where the accelerating mode parameters are given by: $v_g$ is the group velocity, $\phi$ is the synchronous phase advance ($=5\pi/6$), grad the accelerating gradient, and R and Q are the shunt impedance and quality factor evaluated at the synchronous phase. This model implies a quadratic dependence on the group velocity and thus reducing the group velocity will enable the power absorbed to be reduced. Reducing the average iris radius reduces the group velocity but at the same time the short range transverse wakefield is increased and in order to prevent this from occurring we increased the synchronous phase from $2\pi/3$ to $5\pi/6$. The new high phase advance structures maintain the same ratio of average iris radius to free space wavelength ($= a/\lambda = 0.18$) but with a reduced group velocity. The RDDS series had an intial group velocity of 12% of the velocity of light, whereas the latest series of structures have initial group velocities of 3% and 5% of the velocity of light and in order to investigate the occurrence of breakdown as a function of structure length 90cm and 60cm structures are about to be fabricated incorporating detuning of the dipole frequencies of the cells.

Here, we examine the long range wakefield for two structures in which we prescribe an error function variation on the synchronous frequency of each individual cell and provide local damping of each cell with a Q ~1000. The results of such local damping are shown in the following section together with interleaving of the structure frequencies.

## 2. WAKE FUNCTION CALCULATION

We utilize a two-band equivalent circuit model [6] in order to analyse the dispersion characteristics of the first two dipole bands. The computer code $\Omega$2D [7] is used to obtain the zero and $\pi$ modes and the circuit model provides all remaining points. The model described in [6] is a little different from what is required and we remove the manifold and insert a resistor in series with the capacitances in order to model local damping, the subject of this paper. The wake function is obtained from the inverse Fourier transform of the spectrum function [8]. We apply this method to obtain the spectrum function and wake function for H90VG5 and H60VG3 with a cell Q of 1000 and the results are shown in fig 1 and 2 respectively.

*Supported under U.S. DOE contract DE-AC03-76SF00515
†Supported under U.S. DOE grant DE-FG03-93ER40

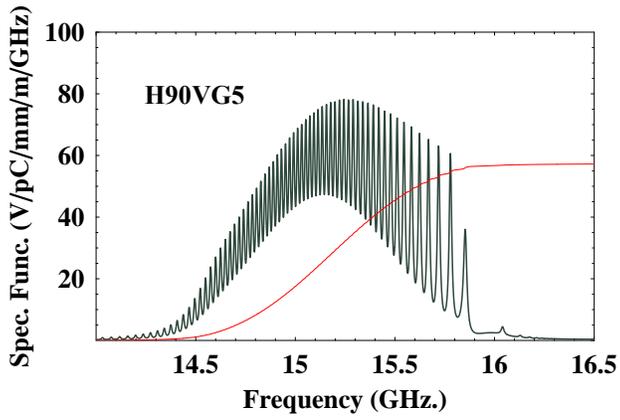
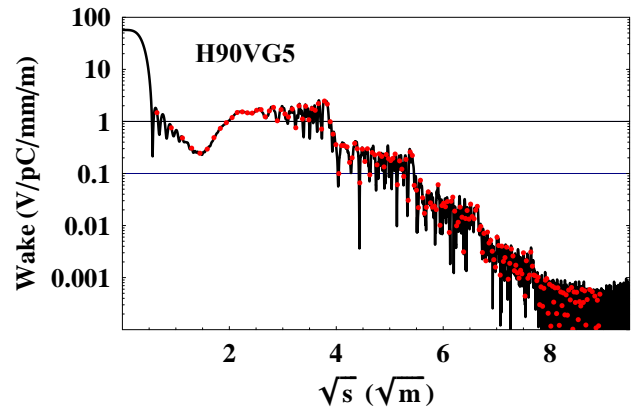
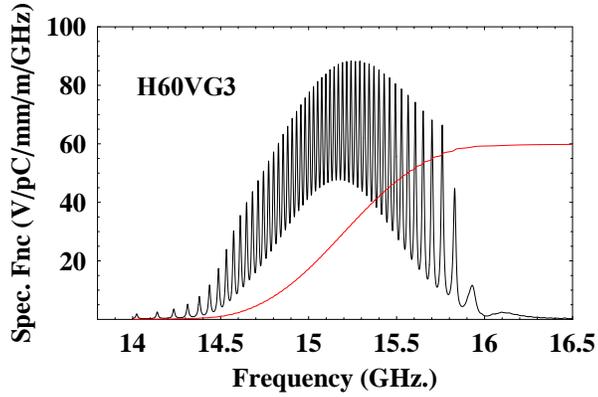
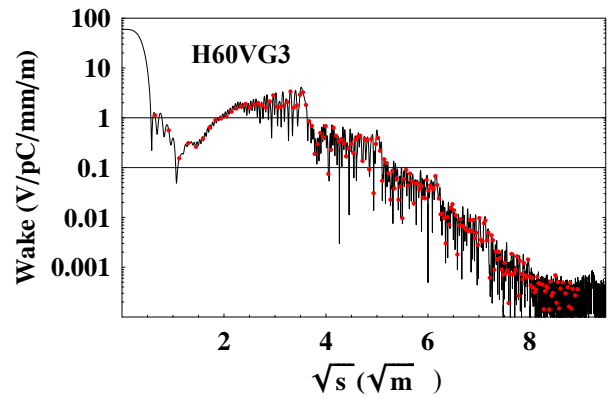
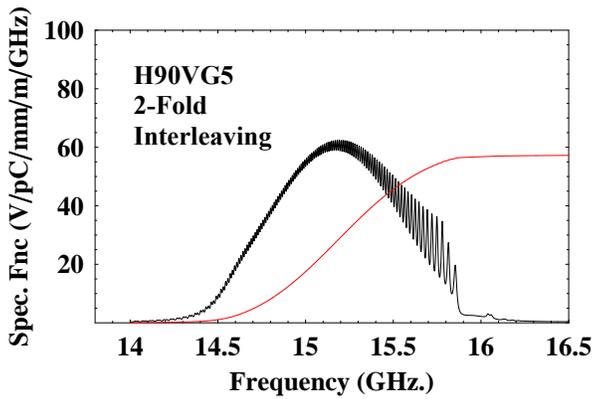
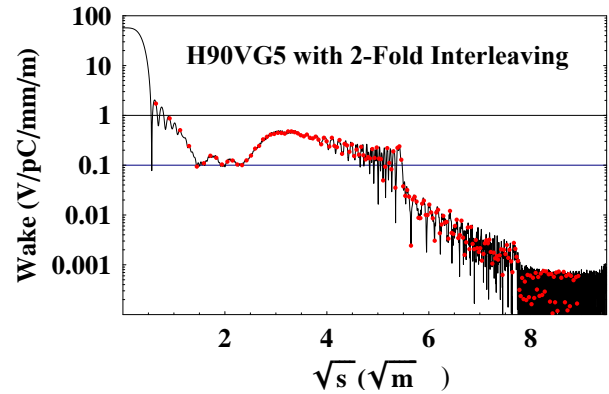
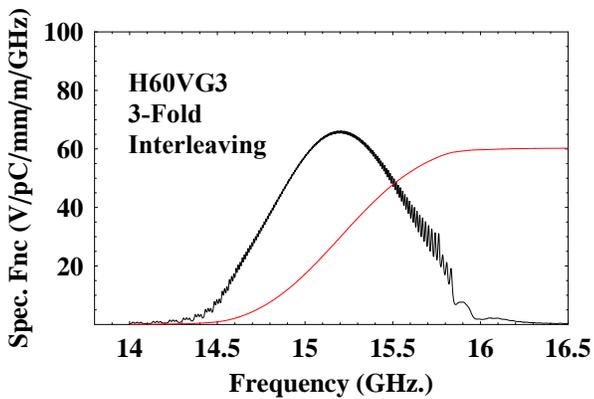
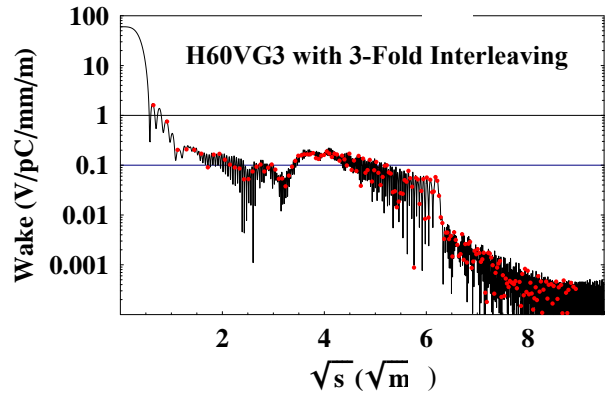

Figure 1: Spectral functions for H90VG5 and H60VG3 including interleaving of the dipole frequencies.

Figure 2: Wake envelope functions for H90VG5 and H60VG3 Including interleaving of the dipole frequencies. (s is the distance behind the leading charge)

Earlier beam dynamics simulations have indicated that the wake field at the bunch locations must be below 1V/pC/mm/m in order that BBU not occur. Each bunch (in the present JLC/NLC design) is spaced from its neighbour by 42cm and it is clear from fig. 2 that there are many positions in which the wakefield is greater than 1 V/pC/mm/m. Thus, either heavy Q loading must be utilized to damp down the wakefield or the dipole frequencies of one particular structure must be interleaved with those from other structures. We adopt this method of interleaving for H90VG5 in which we take the original error function distribution in the cell synchronous frequencies and double the number of frequencies in the given distribution, picking the odd values for one structure and the even values for the neighbouring accelerating structure and from this we end up with 2 structures on a girder (of total length 1.8m). A similar technique is applied to H60VG3. The interleaved spectral function for H90VG5 is significantly improved from a single structure. However, the wakefield for an interleaved version H60VG3 shows superior damping and it easily meets the beam dynamics requirements.

## 3. BEAM DYNAMICS

The sum wakefield (defined at a particular bunch as the sum of the wakefield from all the bunches ahead of that bunch) provides a measure as to whether or not BBU will occur along the linac. BBU is indeed a complex phenomena but through extensive tracking simulations we have found that, provided the standard deviation of the sum wakefield from the mean value ($S_\sigma$) is less than 1 V/pC/mm/m then BBU will not occur. In fig 3 $S_\sigma$ has been calculated for the both interleaved structures as a function of a small fractional change in the bunch spacing ($\Delta s_b$, corresponding to a systematic error in the cell frequencies). H60VG3 clearly remains well below unity and tracking the beam (initially offset by 1µm) through 5,000 or more structures (shown in fig 4) indicates that little emittance growth occurs even at the worst case value of $S_\sigma = 0.69$.

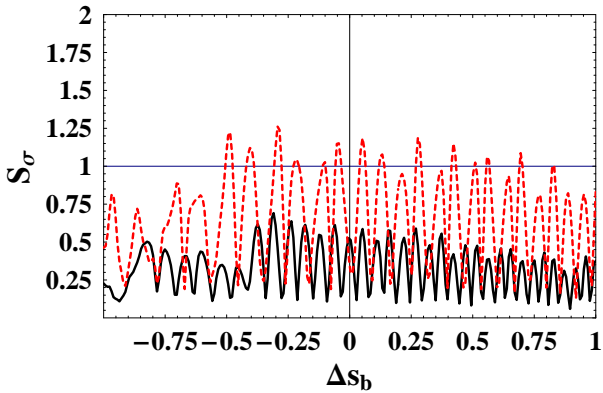

Fig. 3: Standard deviation of sum wake function from the mean value for H90VG5 with two-fold interleaving (shown dashed in red) compared to H60VG3 with three-fold interleaving (solid black)

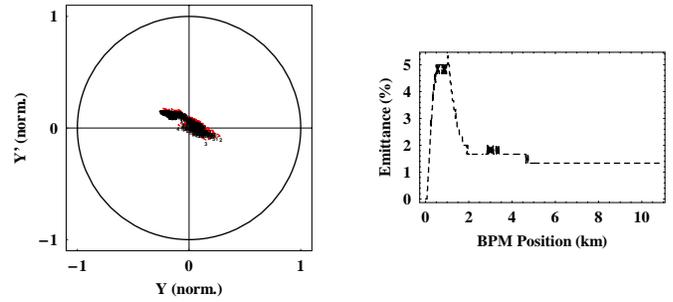

Fig. 4: Phase space and emittance growth for H60VG3 with three-fold interleaving (maximum of $S_\sigma = 0.69$ at a 0.31% reduction in the nominal bunch spacing)

The two-fold interleaved version of H90VG5 does not show a significant emittance dilution unless we take the worst-case value of $\Delta s_b = -0.293\%$ and $S_\sigma =1.26$ then we find that BBU does indeed occur (see fig 5 in which the emittance has grown by 80%). However, in practice the systematic error is unlikely to be unaccompanied by random errors in the cell frequencies from structure to structure and these have been shown to minimise stop to reduce the effective wakefield and to stop BBU from occurring [9].

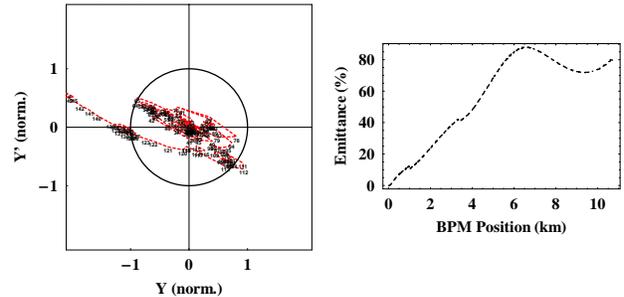

Fig. 5: Phase space and emittance growth for H90VG5 with 2-fold interleaving of structures(maximum of $S_\sigma$ =1.26 at 0.293% reduction in the nominal bunch spacing)

## 4. CONCLUSIONS

The wakefield in the new low group velocity short structures, H90VG5 and H60VG3 meet the beam dynamics requirements of giving rise to no more than 3% emittance dilution and BBU does not occur provided the frequencies of the structures are interleaved.